\documentclass[runningheads]{llncs}
\usepackage[T1]{fontenc}
\usepackage{graphicx,verbatim}
\usepackage{booktabs}
\usepackage{multirow}
\usepackage{makecell}
\usepackage{cite}

\usepackage{amsmath} 
\usepackage{multirow}
\begin{document}
\title{Pathology-Informed Latent Diffusion Model for Anomaly Detection in Lymph Node Metastasis}
%
\begin{comment}  %% Removed for anonymized MICCAI 2025 submission
\author{First Author\inst{1}\orcidID{0000-1111-2222-3333} \and
Second Author\inst{2,3}\orcidID{1111-2222-3333-4444} \and
Third Author\inst{3}\orcidID{2222--3333-4444-5555}}
%
\authorrunning{F. Author et al.}
\institute{Princeton University, Princeton NJ 08544, USA \and
Springer Heidelberg, Tiergartenstr. 17, 69121 Heidelberg, Germany
\email{lncs@springer.com}\\
\url{http://www.springer.com/gp/computer-science/lncs} \and
ABC Institute, Rupert-Karls-University Heidelberg, Heidelberg, Germany\\
\email{\{abc,lncs\}@uni-heidelberg.de}}

\end{comment}
\author{Jiamu Wang\inst{1} \and
Keunho Byeon\inst{1} \and 
Jinsol Song\inst{1} \and 
Anh Nguyen\inst{1} \and 
Sangjeong Ahn\inst{2} \and 
Sung Hak Lee\inst{3}$^*$ \and 
Jin Tae Kwak\inst{1}$^*$
}

% index{Wang, Jiamu}
% index{Byeon, Keunho}
% index{Song, Jinsol}
% index{Nguyen, Anh}
% index{Ahn, Sangjeong}
% index{Lee, Sung Hak}
% index{Kwak, Jin Tae}
\authorrunning{J. Wang et al.}

\institute{
School of Electrical Engineering, Korea University, Seoul 02841, Korea\\
\email{\{taurusmumu, bkh5922, truetg, ngtienanh, jkwak\}@korea.ac.kr}
\and
Department of Pathology, Korea University Anam Hospital and Department of Biomedical Informatics, Korea University College of Medicine, Seoul 02841, Korea\\
\email{vanitas80@korea.ac.kr}
\and
Department of Hospital Pathology, Seoul St. Mary’s Hospital, College of Medicine, The Catholic University of Korea, Seoul 06591, Korea\\
\email{hakjjang@catholic.ac.kr}
}

% \author{Anonymized Authors}  %% Added for anonymized MICCAI 2025 submission
% \authorrunning{Anonymized Author et al.}
% \institute{Anonymized Affiliations \\
%     \email{email@anonymized.com}}

\maketitle              \def\thefootnote{*}\footnotetext{These authors contributed equally to this work}
\begin{abstract}
Anomaly detection is an emerging approach in digital pathology for its ability to efficiently and effectively utilize data for disease diagnosis. While supervised learning approaches deliver high accuracy, they rely on extensively annotated datasets, suffering from data scarcity in digital pathology. Unsupervised anomaly detection, however, offers a viable alternative by identifying deviations from normal tissue distributions without requiring exhaustive annotations. Recently, denoising diffusion probabilistic models have gained popularity in unsupervised anomaly detection, achieving promising performance in both natural and medical imaging datasets. Building on this, we incorporate a vision-language model with a diffusion model for unsupervised anomaly detection in digital pathology, utilizing histopathology prompts during reconstruction. Our approach employs a set of pathology-related keywords associated with normal tissues to guide the reconstruction process, facilitating the differentiation between normal and abnormal tissues. To evaluate the effectiveness of the proposed method, we conduct experiments on a gastric lymph node dataset from a local hospital and assess its generalization ability under domain shift using a public breast lymph node dataset. The experimental results highlight the potential of the proposed method for unsupervised anomaly detection across various organs in digital pathology. Code: https://github.com/QuIIL/AnoPILaD.

% Authors must provide keywords and are not allowed to remove this Keyword section.

\end{abstract}
\section{Introduction}

Lymph node metastasis is a crucial prognostic factor in cancer progression and treatment decisions~\cite{metastasis}. With the advent of digital pathology, several artificial intelligence approaches have been proposed to automate the detection of lymph node metastasis within tissues. Many of these methods are based on supervised learning methods, leveraging convolutional neural networks (CNNs) and transformer-based models to identify metastasis with high accuracy~\cite{digital_deeplearning}. While effective, these methods heavily depend on exhaustive expert annotations, which are time-consuming and resource-intensive. To address this limitation, unsupervised learning has emerged as a viable alternative, as it does not demand manual annotations. In unsupervised learning, a model is trained solely on normal (in-distribution) samples to learn a representation of in-distribution patterns. The trained model then detects abnormal or out-of-distribution (OOD) samples by identifying deviations from the learned in-distribution patterns, making it particularly well-suited for large-scale applications in digital pathology, where annotated abnormal samples are scarce~\cite{ood_survey}.

Recently, generative models have been widely used for unsupervised anomaly detection, with two popular approaches: density-based methods and reconstruction-based methods. Density-based methods, such as variational autoencoders (VAEs) ~\cite{vae, lr, llr, input_cplex}, learn representations of in-distribution data, assigning higher likelihoods to in-distribution samples and lower likelihoods to OOD samples. In contrast, reconstruction-based methods are trained exclusively on normal data to guarantee poor reconstruction quality for abnormal samples and high reconstruction quality for normal samples. These include autoencoder (AE)-based models~\cite{autoencoder, stae, memae}, Generative Adversarial Network (GAN)~\cite{gan, anogan, fanogan}-based models and denoising diffusion probabilistic model (DDPM)-based~\cite{ddpm, anoddpm, ddpm_ood, anoddpm_lymphnode} models. 

In this paper, we introduce AnoPILaD, a Pathology-Informed Latent Diffusion model for anomaly detection in lymph node pathology images. This framework combines a latent diffusion model (LDM)~\cite{ldm} and a vision-language model (VLM)~\cite{vlm} for an improved identification of anomalies in pathology images. AnoPILaD utilizes a LDM to learn a compact representation of normal images in a latent space via iterative diffusion and denoising processes while preserving critical histopathological features. AnoPILaD also adopts a VLM to select pathology-specific normal keywords, semantically guiding the reconstruction process towards a specific direction. In this manner, AnoPILaD achieves small deviations for normal samples and large deviations for abnormal samples, enhancing the accuracy and robustness of anomaly detection.

\section{Methods}
\subsection{Pathology-Informed Latent Diffusion Model}
Recent research often employs DDPMs for reconstruction-based anomaly detection. A trained DDPM $p_{\theta}$ generates samples that match the in-distribution patterns $\mathbf{z_0}\sim q(\mathbf{z_0})$ by adding noise in a forward process (diffusion process), which has tractable posteriors at time $t-1$:
\begin{equation}
    q(\mathbf{z}_{t-1} |\mathbf{z}_{t}, \mathbf{z}_0) = \mathcal{N} \left( \mathbf{z}_{t-1} \middle| \tilde{\mu} (\mathbf{z}_t(\mathbf{z_0, \mathbf{\epsilon}}), t), \tilde{\beta}_t \mathbf{I} \right)
\end{equation}

\noindent where $t\sim[1,1000]$, $\epsilon \sim \mathcal{N}(0,\mathbf{I})$, $\mathbf{z_t}$ is a noisy sample, and $\tilde{\beta}_t$ is a predefined constant. The model progressively removes noise in a reverse process:
\begin{equation}
    p_{\theta}(\mathbf{z}_{t-1} | \mathbf{z}_t) = \mathcal{N} \left( \mathbf{z}_{t-1} \middle| \mu_{\theta} (\mathbf{z}_t(\mathbf{z_0, \mathbf{\epsilon_{\theta}}}), t), \tilde{\beta}_t \mathbf{I} \right)
\end{equation}

\noindent where $\mathbf{\epsilon_{\theta}}$ is an approximator to predict $\epsilon$ from $\mathbf{z_t}$. The model is trained by decreasing the KL divergence between two Gaussians, and a simple objective function is given by:
\begin{equation}
\mathcal{L} = E_{\mathbf{z_t},t,\epsilon \sim \mathcal{N}(0,1)} \left[ \left\| \epsilon - \epsilon_\theta(\mathbf{z_t}, t) \right\|_2^2 \right].
\end{equation}

\begin{figure}
\includegraphics[width=\textwidth]{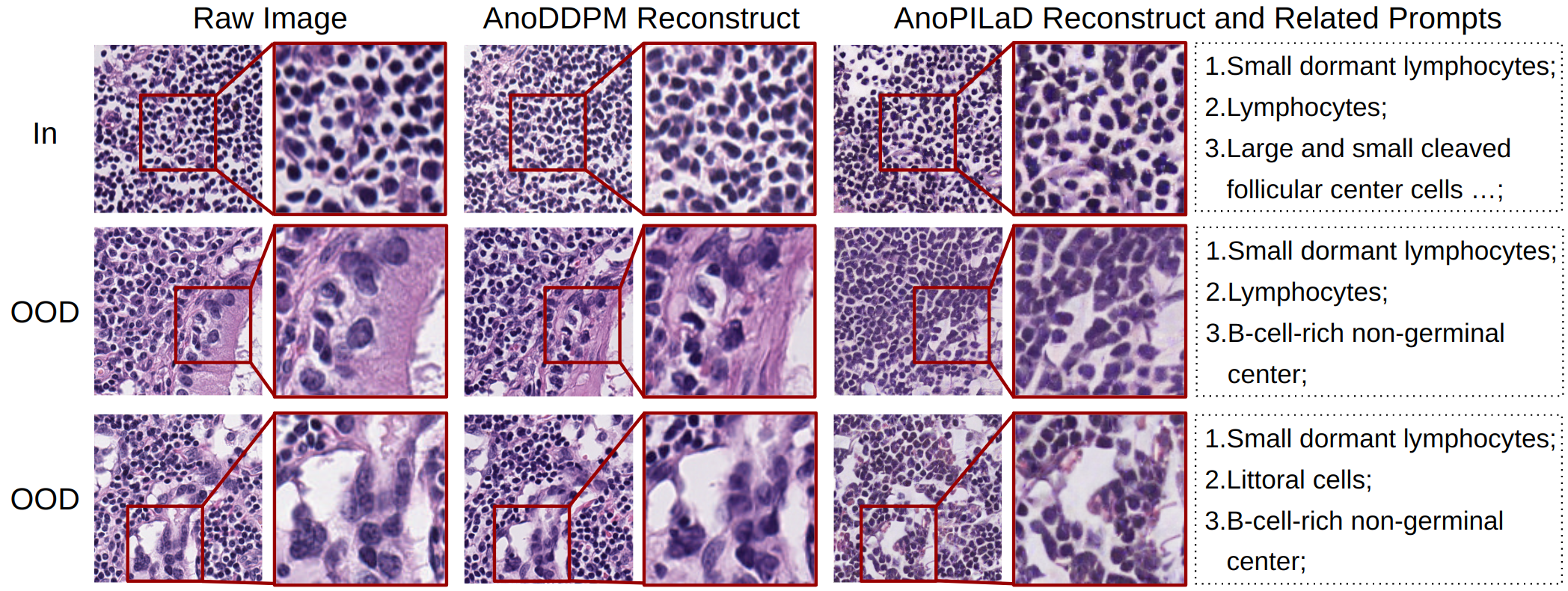}
\caption{Lymph node image reconstruction using diffusion-based methods. Using text prompts, AnoPILaD produces normal-like pathology images for both in-distribution (in) and out-of-distribution (OOD) samples.} \label{fig1}
\end{figure}

\noindent A previous work~\cite{anoddpm_lymphnode} adopted a DDPM for detecting breast lymph node metastasis, referred to as AnoDDPM, where normal samples are in-distribution data and metastasis samples are OOD data. The pretrained DDPM denoised partially diffused inputs $\mathbf{z_t}\sim q(\mathbf{z_t}|\mathbf{z_0})$ while steering the reverse process toward in-distribution patterns and get the reconstruction $\hat{\mathbf{z_0}}\sim p_\theta(\mathbf{z_0}|\mathbf{z_{1:t}})$. Then, the discrepancy between the input $\mathbf{z_0}$ and its reconstruction $\hat{\mathbf{z_0}}$ serves as the anomaly score, which is expected to be small for normal samples and large for metastasis samples. Though successful, AnoDDPM exhibited a substantial number of false positive~\cite{anoddpm_lymphnode}, indicating its insufficient ability to differentiate samples from normal and OOD distributions (Fig. \ref{fig1}). To address this issue and improve anomaly detection performance, AnoPILaD integrates a LDM with pathology-specific textual prompts, based on the assumption that these prompts enhance reconstruction quality and, in turn, magnify the contrast between normal and abnormal samples. We train a LDM with the following objective:
\begin{equation}
\mathcal{L} = E_{\mathbf{z_t},t,c,\epsilon \sim \mathcal{N}(0,1)} \left[ \left\| \epsilon - \epsilon_\theta(\mathbf{z_t}, t, c) \right\|_2^2 \right]
\end{equation}
\noindent where $c$ denotes a textual prompt. We conduct pathology-informed reconstruction for lymph node pathology images by using the same reconstruction procedure with AnoDDPM. Only exception is that the reverse process is guided by the textual condition, which is given by $\hat{\mathbf{z_0}}\sim p_\theta(\mathbf{z_0}|\mathbf{z_{1:t}}, c)$.

\subsection{Weighted Prompts Generation}
To introduce a stronger inductive bias to the reconstruction process, we propose to exploit prior pathology knowledge of normal lymph nodes. Specifically, we collect a pool of 74 pathology keywords from the literature, describing characteristics of cells and microenvironments of normal lymph nodes and tissues. These keywords are reviewed and validated by an experienced pathologist to ensure clinical relevance and accuracy. 

\begin{figure}
\includegraphics[width=\textwidth]{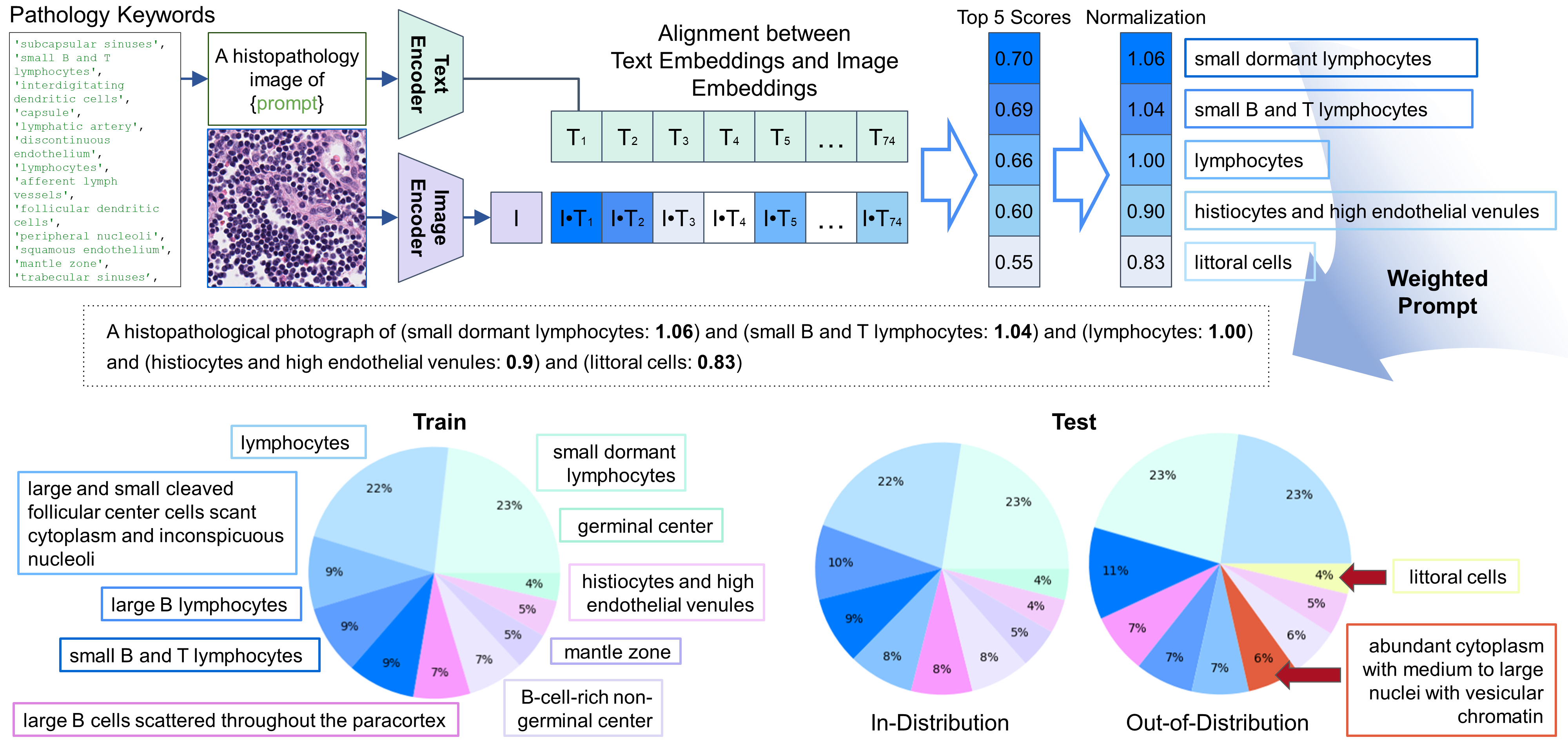}
\caption{(Top) Illustration of generating weighted text prompts. (Bottom) Distribution of top-10 frequent pathology keywords in the training and test sets the local hospital dataset.} \label{fig2}
\end{figure}

Given a pathology image, we align it with the pathology keywords, identify the most relevant keywords, and use them to generate a text prompt, guiding the reconstruction process towards a pathology-informed direction (Fig. \ref{fig2}). To achieve the alignment between pathology images and keywords, we adopt CONCH~\cite{conch}, a vision-language foundation model pre-trained on over 1.17 million pathology image-caption pairs. The image encoder and text encoder of CONCH are used to produce image and text embeddings for each pair of input image and keywords, respectively. Then, we compute cosine similarity scores between image embeddings and text embeddings and choose top-five most similar keywords. The similarity scores of the selected keywords are subsequently normalized by dividing by the median score. Using the selected keywords and their normalized scores, we generate a weighted prompt as shown in Fig. ~\ref{fig2}. The weighted prompt is transformed into an embedding vector, which is fed into the LDM, following the procedure illustrated in the Compel library~\footnote{https://github.com/damian0815/compel}.

\begin{table}[htbp]
\centering
\caption{Two Datasets used in this Study}
\label{tab1}
\resizebox{\linewidth}{!}{
\begin{tabular}{l c c c c c c c c c}
\hline 
\multirow{2}{*}{Dataset} & \multicolumn{4}{c}{\textbf{WSI-level dataset}} & & \multicolumn{4}{c}{\textbf{Patch-level dataset}}\\ %\hline
\cline{2-5} \cline{7-10}
 & $D_{\text{tr}}^{w}$ & $D_{\text{val}}^{w}$ & $D_{\text{in}}^{w}$ & $D_{\text{out}}^{w}$ ($D_{\text{out,m}}^{w}$) & & $D_{\text{tr}}^{p}$ & $D_{\text{val}}^{p}$ & $D_{\text{in}}^{p}$ & $D_{\text{out}}^{p}$ ($D_{\text{out,m}}^{p}$)
\\ \hline
LH & 643 & 50 & 58 & 57 & & 1,373,475 & 102,240 & 174,703 & 115,330
\\ %\hline

C16 & - & 30 & 88 & 44 (20) & & - & 37,056 & 240,139 & 55,659 (55,536)\\ \hline
\end{tabular}
}
\end{table}

\section{Experiment}
\subsection{Dataset}
We obtained a gastric lymph node dataset from two local hospital (LH) sites containing 808 Whole Slide Images (WSIs), where 751 of them are normal and 57 are metastasis WSIs with partial annotation. All the metastasis WSIs are used as an OOD test set ($D^{LH,w}_{out}$). The normal WSIs are split into a training set of 643 WSIs ($D^{LH,w}_{tr}$), a valid set of 50 WSIs ($D^{LH,w}_{val}$), and an in-distribution test set of 58 WSIs ($D^{LH,w}_{in}$). 
We further divide WSIs into patches, resulting in $D^{LH,p}_{tr}$ (1,373,475), $D^{LH,p}_{val}$ (102,240), $D^{LH,p}_{in}$ (174,703 patches, 138054 from $D^{LH,w}_{in}$ and 36649 from normal annotated $D^{LH,w}_{out}$), and $D^{LH,p}_{out}$ (115,330 with fully metastasis annotation). 
Moreover, we employed the Camelyon16 Challenge dataset (C16)~\cite{c16}, a breast lymph node dataset, for independent testing. We utilize 32 normal WSIs as a valid set ($D^{C16,w}_{val}$), 88 normal WSIs as an in-distribution test set ($D^{C16,w}_{in}$), and 40 metastasis WSIs as an OOD test set ($D^{C16,w}_{out}$). Among the 40, 22 WSIs contain tumor cell cluster larger than 2 mm in diameter, designated as C16 Macro ($D^{C16,w}_{\text{out,m}}$). All tumor regions in LH are larger than 2 mm in diameter.
We further divide WSIs into patches, resulting in $D^{C16,p}_{val}$ (37,056), $D^{C16,p}_{in}$ (240,139), and $D^{C16,p}_{out}$ (55,659 patches with fully metastasis annotation). 
For patch-level evaluation, WSIs from both LH and C16 are divided into 256$\times$256 pixel patches using the pixel-level annotations. All WSIs are processed at 20$\times$ magnification.
An overview of the datasets is presented in Table \ref{tab1}.

\subsection{Experiment Details}
To implement AnoPILaD, we utilized the stable diffusion model v1.5. The diffusion model was fine-tuned for 400,000 steps with Adam optimizer, a learning rate of 1e-5, and a batch size of 64 using low-rank adaptation (LORA) with the update matrices dimension of 4. The input image size was 256$\times$256 pixels.
We compared AnoPILaD with both density- and reconstruction-based OOD methods. For density-based methods, we used the negative likelihood (NLL) of a VAE backbone~\cite{vae} and its three variants: Regret~\cite{lr}, LLR~\cite{llr}, and complexity~\cite{input_cplex}. The latent vector size was set to 100, and 64$\times$64 pixels input images were randomly cropped. 
As for reconstruction-based methods, we employed f-AnoGAN~\cite{fanogan}, AE, and MemAE~\cite{memae}. f-AnoGAN used randomly cropped 64$\times$64 input image patches. For AE and MemAE, they follow the architectural design of~\cite{memae} and input image size was 256$\times$256 pixels. 
Both AnoDDPM and AnoPILaD were implemented using Diffusers library~\cite{diffusers} and used a PLMS sampler~\cite{plms} with 100 timesteps in inference. 
Unless otherwise specified, the training procedures followed those outlined in the original work.

%we train a three-layer AE as in []. Memory size is 500 for MemAE and both methods use MSE as the anomaly score and the input image size is 256$\times$256.

%These models were trained for 10 epochs with a batch size of 64, Adam optimizer, and a learning rate of 5e-4. 
% For the LLR method, we trained the background encoder and generator with a perturbation ratio of 0.2, as specified to be optimal in their paper. During testing, we fine-tuned the encoder for each test sample over 100 iterations.
%train a WGAN for fanogan [] with gradient penalty for 10 epochs, followed by 5 epochs with frozen generator and discriminator parameters, using a batch size of 64. The model structure follows the original paper, with a z-space dimension of 128. We use Adam for WGAN training and RMSprop for the encoder, with a learning rate of 5e-5. 

%We use Diffusers library [] to build two diffusion model-based methods, pretrained CelebA-HQ for AnoDDPM and stable 1.5 for AnoPILaD. We fine-tune AnoDDPM for 1M iterations with a batch size of 28 and AnoPILaD for 400,000 steps with a batch size of 64. In addition, we use LORA with the update matrices dimension of 4 to train AnoPILaD. Both models utilize the Adam optimizer with a learning rate of 2e-5 and input image size of 256$\times$256. At test time, we used a PLMS sampler with 100 timesteps.

For each method, we calculated z-scores of the anomaly scores for all the patches in each testing set and evaluated patch-level OOD detection by computing area under a receiver operating characteristic curve (AUC) and area under the precision-recall curve (AUPR). For WSI-level evaluation, we produced a z-score heatmap and applied a morphological erosion operation~\cite{scipy} with a 2$\times$2 window since metastasis area can be very samll. We then assessed the classification and segmentation performance of each model.
For classification, we adopted two methods: maximum z-score ($Z_{MAX}$) and average of 99th percentile z-scores ($Z_{99}$). These are used to calculate WSI-level AUC and AUPR. For segmentation, we calculated the mean patch-level DICE and intersection-over-union (IoU) of $D_{\text{out}}^{w}$ to assess the overlap between the predictions and annotations, and the mean patch-level true negative rate (TNR) of $D_{\text{in}}^{w}$ since there is no positive regions. The segmentation prediction threshold was zero.
To decide reconstruction timestep for two diffusion model-based methods, we tested eight timesteps values as in~\cite{anoddpm_lymphnode} and choose the timestep according to the best WSI classification performance on LH test set, which is 674 for both methods.

\section{Results}
We trained AnoPILaD and all competing models with only $D_{tr}$ and separately evaluated their patch-level and WSI-level performance on two distinct datasets originated from different organs. Thus, this evaluation provides insights into the robustness of the models against domain shifts due to variations in tissue types.

\begin{table}[htbp]
\centering
\caption{Patch-level Anomaly Detection Results}
\label{tab3}
\resizebox{\linewidth}{!}{
\begin{tabular}{llccccccccc}
\hline
%\multicolumn{10}{c}{Local Hospital Dataset}
%\\ \hline
 & & NLL & Regret & LLR & complexity & f-AnoGAN & AE & MemAE & AnoDDPM & AnoPILaD
\\ \hline
\multirow{2}{*}{LH} & AUC  &  0.4982 & 0.6720 & 0.6078 & 0.7931 & 0.2289 & 0.9254 & 0.9290 & 0.8555 & $\mathbf{0.9587}$
\\ 
 & AUPR &  0.5552 & 0.6718 & 0.6260 & 0.7139 & 0.3377 & 0.8906 & 0.8886 & 0.7841 & $\mathbf{0.9499}$
\\ \hline
%\multicolumn{10}{c}{C16 Dataset}
%\\ \hline
\multirow{2}{*}{C16} & AUC  &  0.4983 & 0.6720 & 0.7065 & 0.7752 & 0.1735 & 0.6584 & 0.6611 & 0.6857 & $\mathbf{0.8884}$
\\ 
 & AUPR &  0.5552 & 0.6718 & 0.4765 & 0.5140 & 0.1104 & 0.4759 & 0.4880 & 0.5741 & $\mathbf{0.6987}$
\\ \hline
\end{tabular}
}
\end{table}

Table~\ref{tab3} presents the performance of patch-level anomaly detection. AnoPILaD substantially outperformed all other methods.
Among the four density-based methods (NLL, Regret, LLR, and complexity), NLL failed to distinguish OOD patches from normal patches. Its variants improved performance, with the complexity method achieving the largest improvement. Despite these improvements, complexity remained substantially inferior to AnoPILaD, with a large gap of $\sim$0.16 AUC and $\sim$0.23 AUPR.
Among the reconstruction-based methods, f-AnoGAN demonstrated the poorest performance, while AE-based methods (AE and MemAE) achieved the highest scores. In LH, their AUC and AUPR scores were approximately 0.03 and 0.05 lower than AnoPILaD, respectively. However, in C16, the performance gap substantially increased to $\sim$0.22 in AUC and $\sim$0.22 in AUPR, indicating the superior robustness of AnoPILaD against domain shifts due to differences in organ types.

\begin{table}[htbp]
\centering
\caption{WSI-level Anomaly Detection Results}
\label{tab4}
\resizebox{\linewidth}{!}{
% \begin{tabular}{l|c|c|c|c|c|c|c|c|c|c|c|c}
\begin{tabular}{lll   ccc c  ccc c  ccc}
\hline
& & & \multicolumn{3}{c}{LH} & & \multicolumn{3}{c}{C16} & & \multicolumn{3}{c}{C16 Macro} 
\\ \cline{4-6} \cline{8-10} \cline{12-14} %\hline
\multicolumn{2}{l}{\textit{Classification}} && AUC & AUPR & - && AUC & AUPR & - && AUC & AUPR & - 
% \\ \cline{2-10}

\\ \hline
\multirow{2}{*}{AE} & $Z_{max}$              && 0.9622 & 0.9612 & - && 0.6612 & 0.5961 & - && 0.6398 & 0.3677 & -  
                    % & \multirow{2}{*}{0.7981} & \multirow{2}{*}{0.3812} & \multirow{2}{*}{0.2902}
                    % & \multirow{2}{*}{0.4536} & \multirow{2}{*}{0.1745} & \multirow{2}{*}{0.1317}
                    % & \multirow{2}{*}{0.4536} & \multirow{2}{*}{0.3249} & \multirow{2}{*}{0.2549}
\\ %\hline
 & $Z_{99}$               && 0.9395 & 0.9381 & - && 0.5798 & 0.5217 & - && 0.5523 & 0.2918 & -  
\\ \hline
\multirow{2}{*}{MemAE} & $Z_{max}$           && 0.9504 & 0.9440 & - && 0.6505 & 0.5689 & - && 0.6381 & 0.3657 & -
                    % & \multirow{2}{*}{0.7957} & \multirow{2}{*}{0.3863} & \multirow{2}{*}{0.2932}
                    % & \multirow{2}{*}{0.5593} & \multirow{2}{*}{0.1377} & \multirow{2}{*}{0.1039}
                    % & \multirow{2}{*}{0.5594} & \multirow{2}{*}{0.2173} & \multirow{2}{*}{0.2124}
\\ %\hline
 & $Z_{99}$            && 0.9365 & 0.9382 & - && 0.5686 & 0.5313 & - && 0.5597 & 0.3141 & - 
\\ \hline
\multirow{2}{*}{AnoDDPM} & $Z_{max}$         && 0.7840 & 0.6616 & - && 0.4992 & 0.3885 & - && 0.4926 & 0.2146 & -  
                    % & \multirow{2}{*}{0.7850} & \multirow{2}{*}{0.4319} & \multirow{2}{*}{0.3259}
                    % & \multirow{2}{*}{0.7842} & \multirow{2}{*}{0.1765} & \multirow{2}{*}{0.1142}
                    % & \multirow{2}{*}{0.7842} & \multirow{2}{*}{0.3131} & \multirow{2}{*}{0.2075}
\\ %\hline
 &  $Z_{99}$          && 0.9383 & 0.8995 & - && 0.4551 & 0.3905 & - && 0.5119 & 0.2347 & -
\\ \hline
\multirow{2}{*}{AnoPILaD} & $Z_{max}$        && 0.9837 & 0.9740 & - &&$\mathbf{0.6745}$ & $\mathbf{0.6140}$ & -& &$\mathbf{0.8062}$ & $\mathbf{0.5965}$ & - 
                    % & \multirow{2}{*}{$\mathbf{0.8097}$} & \multirow{2}{*}{$\mathbf{0.4322}$} & \multirow{2}{*}{$\mathbf{0.3311}$}
                    % & \multirow{2}{*}{$\mathbf{0.8312}$} & \multirow{2}{*}{$\mathbf{0.3098}$} & \multirow{2}{*}{$\mathbf{0.2326}$}
                    % & \multirow{2}{*}{$\mathbf{0.8312}$} & \multirow{2}{*}{$\mathbf{0.5420}$} & \multirow{2}{*}{$\mathbf{0.4275}$}
\\ %\hline
 & $Z_{99}$         && $\mathbf{0.9943}$ & $\mathbf{0.9948}$ & - && 0.6367 & 0.5902 & - && 0.8023 & 0.5886 & - 
\\ \hline
%\\ \hline
% & \multicolumn{9}{c}{Segmentation} \\ \hline
& & & \multicolumn{3}{c}{LH} & & \multicolumn{3}{c}{C16} & & \multicolumn{3}{c}{C16 Macro} 
\\ \cline{4-6} \cline{8-10} \cline{12-14} %\hline
\multicolumn{2}{l}{\textit{Segmentation}} && TNR & DICE & IoU && TNR & DICE & IoU && TNR & DICE & IoU  \\ \hline
AE        &          && 0.7981 & 0.3812 & 0.2902 && 0.4536 & 0.1745 & 0.1317 && 0.4536 & 0.3249 & 0.2549  \\ %\hline
MemAE     &         && 0.7957 & 0.3863 & 0.2932 && 0.5593 & 0.1377 & 0.1039 && 0.5594 & 0.2173 & 0.2124  \\ %\hline
AnoDDPM   &          && 0.7850 & 0.4319 & 0.3259 & &0.7842 & 0.1765 & 0.1142 && 0.7842 & 0.3131 & 0.2075  \\ %\hline
AnoPILaD  &         && $\mathbf{0.8097}$ & $\mathbf{0.4322}$ & $\mathbf{0.3311}$ 
                    && $\mathbf{0.8312}$ & $\mathbf{0.3098}$ & $\mathbf{0.2326}$ 
                    && $\mathbf{0.8312}$ & $\mathbf{0.5420}$ & $\mathbf{0.4275}$ \\ \hline %\hline

\end{tabular}
}
\end{table}

At the WSI-level, we further confirmed our findings from patch-level anomaly detection. Table~\ref{tab4} shows the WSI-level anomaly detection results for the four top-performing models (AE, MemAE, AnoDDPM, and AnoPILaD) in patch-level anomaly detection. For the classification of normal and abnormal slides, AnoPILaD was superior to all competing models across both datasets, evaluation metrics, and scoring strategies ($Z_{MAX}$ and $Z_{99}$). The performance of AnoPILaD and other three models substantially varied between LH and C16. In LH, the four models obtained 0.7840$\sim$0.9943 AUC and 0.6616$\sim$0.9948 AUPR, whereas their performance dropped in C16, with 0.4551$\sim$0.6745 AUC and 0.3885$\sim$0.6140 AUPR. Nonetheless, AnoPILaD exhibited the smallest best performance drop of 0.3092$\sim$0.3576 AUC and 0.3600$\sim$0.4046 AUPR, highlight its robustness across datasets and organ types. In regard to large metastatic regions (C16 Macro), the performance generally declined for all models; however, AnoPILaD obtained AUCs ranging from 0.8023 to 0.8062, suggesting its superior potential for cross-organ anomaly detection in lymph nodes.

\begin{figure}
\includegraphics[width=\textwidth]{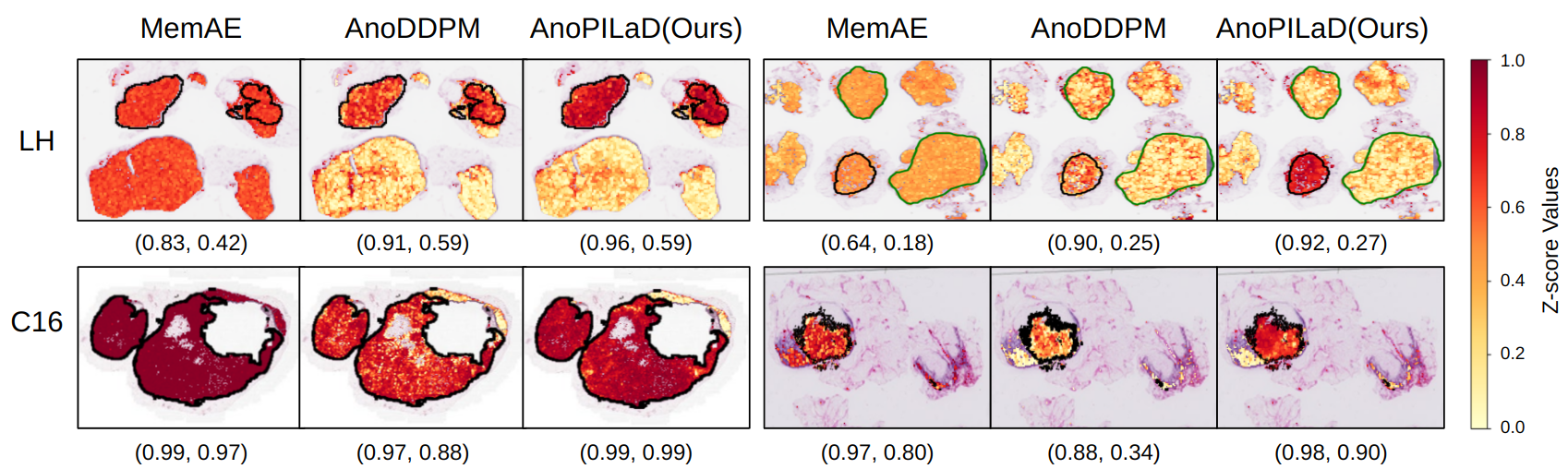}
\caption{Z-score heatmaps for four metastasis slides from two datasets with black contour showing metastasis annotation and green contour showing normal annotations. Numbers in parentheses denotes ($Z_{99}$, Dice) per slide.} \label{fig3}
\end{figure}

Evaluating the segmentation results of metastatic regions, the strength of AnoPILaD was obvious, achieving the highest scores for all evaluation scenarios. We also observed that there are clear differences between the two anomaly detection approaches. Although AE-based methods (AE and MemAE) obtained comparable classification performance with diffusion-based methods (AnoPILaD and AnoDDPM), their segmentation performance was substantially poorer. As shown in Fig~\ref{fig3}, MemAE assigns similar scores for all pixels in a slide, indicating a lack of specificity in detecting metastatic regions. The behavior of AE is almost identical to MemAE. These results suggest the importance of segmentation performance in the evaluation of anomaly detection, as classification alone may not accurately assess a model's ability to localize abnormal regions.

We visualize and compare the reconstruction results by AnoPILaD and AnoDDPM (Fig~\ref{fig1}). 
Starting from raw images, AnoPILaD and AnoDDPM aim to reconstruct normal-like structures with and without textual guidance, respectively. For an in-distribution sample (first row), both models successfully generate images with normal histologic features, preserving small, dense lymphocytic structures. However, as for OOD samples (second and third rows), which contain metastatic regions with disrupted tissue architecture, the differences between two models become more apparent. AnoDDPM partially reconstructs metastatic regions and fails to fully suppress pleomorphic nuclei and fibrotic tissue, leading to residual abnormalities that obscure the boundary between normal and abnormal structures. 
In contrast, with the guidance of text prompts, AnoPILaD generates images with more uniform lymphocytic arrangements, while suppressing distortions in tissue architecture. This results in a clearer distinction between normal and metastatic regions.

\section{Conclusion}
We propose AnoPILaD, a pathology-informed LDM for unsupervised anomaly detection in lymph nodes. By leveraging histological context provided through prompts, AnoPILaD introduces a stronger inductive bias during the reconstruction process, enhancing sensitivity to abnormal features and improving detection performance. Evaluating both patch-level and slide-level performance across two organ types, AnoPILaD substantially outperforms other anomaly detection methods including both density- and reconstruction-based approaches. The future study will entail the extension of AnoPILaD to further enhance its performance on cross-organ datasets, improving its adaptability and robustness in broader pathological applications.

\begin{thebibliography}{8}
\bibitem{lr}
Xiao, Zhisheng, Qing Yan, and Yali Amit. "Likelihood regret: An out-of-distribution detection score for variational auto-encoder." Advances in neural information processing systems 33 (2020): 20685-20696.

\bibitem{anoddpm_lymphnode}
Linmans, Jasper, et al. "Diffusion models for out-of-distribution detection in digital pathology." \textit{Medical Image Analysis} 93 (2024): 103088.

\bibitem{ddpm_ood}
Graham, Mark S., et al. "Denoising diffusion models for out-of-distribution detection." \textit{Proceedings of the IEEE/CVF Conference on Computer Vision and Pattern Recognition}. 2023.

\bibitem{anoddpm}
Wyatt, Julian, et al. "Anoddpm: Anomaly detection with denoising diffusion probabilistic models using simplex noise." \textit{Proceedings of the IEEE/CVF Conference on Computer Vision and Pattern Recognition}. 2022.

\bibitem{memae}
Gong, Dong, et al. "Memorizing normality to detect anomaly: Memory-augmented deep autoencoder for unsupervised anomaly detection." \textit{Proceedings of the IEEE/CVF international conference on computer vision}. 2019.

\bibitem{stae}
Chong, Yong Shean, and Yong Haur Tay. "Abnormal event detection in videos using spatiotemporal autoencoder." \textit{Advances in Neural Networks-ISNN 2017: 14th International Symposium, ISNN 2017, Sapporo, Hakodate, and Muroran, Hokkaido, Japan, June 21–26, 2017, Proceedings, Part II 14}. Springer International Publishing, 2017.

\bibitem{autoencoder}
Hinton, Geoffrey E., and Richard Zemel. "Autoencoders, minimum description length and Helmholtz free energy." \textit{Advances in neural information processing systems} 6 (1993)

\bibitem{metastasis}
Nathanson, S. David. "Insights into the mechanisms of lymph node metastasis." \textit{Cancer} 98.2 (2003): 413-423.

\bibitem{digital_deeplearning}
Budginaite, Elzbieta, et al. "Computational methods for metastasis detection in lymph nodes and characterization of the metastasis-free lymph node microarchitecture: A systematic-narrative hybrid review." \textit{Journal of Pathology Informatics} 15 (2024): 100367.

\bibitem{ood_survey}
Salehi, Mohammadreza, et al. "A unified survey on anomaly, novelty, open-set, and out-of-distribution detection: Solutions and future challenges." arXiv preprint arXiv:2110.14051 (2021).

\bibitem{ldm}
Rombach, Robin, et al. "High-resolution image synthesis with latent diffusion models." \textit{Proceedings of the IEEE/CVF conference on computer vision and pattern recognition}. 2022.

\bibitem{vlm}
Zhang, Jingyi, et al. "Vision-language models for vision tasks: A survey." \textit{IEEE Transactions on Pattern Analysis and Machine Intelligence} (2024).

\bibitem{ddpm}
Ho, Jonathan, Ajay Jain, and Pieter Abbeel. "Denoising diffusion probabilistic models." \textit{Advances in neural information processing systems} 33 (2020): 6840-6851.

\bibitem{conch}
Lu, Ming Y., et al. "Towards a visual-language foundation model for computational pathology." \textit{arXiv preprint arXiv:2307.12914} (2023).

\bibitem{c16}
Bejnordi, Babak Ehteshami, et al. "Diagnostic assessment of deep learning algorithms for detection of lymph node metastases in women with breast cancer." \textit{Jama} 318.22 (2017): 2199-2210.

\bibitem{diffusers}
Diffusers: State-of-the-art diffusion models" by Patrick von Platen, Suraj Patil, Anton Lozhkov, et al. (2022), available on GitHub as a repository named "huggingface/diffusers

\bibitem{plms}
Liu, Luping, et al. "Pseudo numerical methods for diffusion models on manifolds." arXiv preprint arXiv:2202.09778 (2022).

\bibitem{vae}
Kingma, Diederik P. "Auto-encoding variational bayes." \textit{arXiv preprint arXiv:1312.6114} (2013).

\bibitem{input_cplex}
Serrà, Joan, et al. "Input complexity and out-of-distribution detection with likelihood-based generative models." \textit{arXiv preprint arXiv:1909.11480} (2019).

\bibitem{llr}
Ren, Jie, et al. "Likelihood ratios for out-of-distribution detection." \textit{Advances in neural information processing systems} 32 (2019).

\bibitem{fanogan}
Schlegl, Thomas, et al. "f-AnoGAN: Fast unsupervised anomaly detection with generative adversarial networks." \textit{Medical image analysis} 54 (2019): 30-44.

\bibitem{anogan}
Schlegl, Thomas, et al. "Unsupervised anomaly detection with generative adversarial networks to guide marker discovery." \textit{International conference on information processing in medical imaging}. Cham: Springer International Publishing, 2017.

\bibitem{gan}
Goodfellow, Ian, et al. "Generative adversarial networks." \textit{Communications of the ACM} 63.11 (2020): 139-144.

\bibitem{scipy}
Virtanen, Pauli, et al. "SciPy 1.0: fundamental algorithms for scientific computing in Python." Nature methods 17.3 (2020): 261-272.
\end{thebibliography}
%

\end{document}